\documentclass[apjl]{emulateapj}
\usepackage{graphicx}
\usepackage{amsmath,amssymb}
\usepackage{bm}
\usepackage{mathtools}
\usepackage{txfonts}
\usepackage{natbib}
\usepackage{multirow}
\usepackage{color}
\usepackage[colorlinks=true,citecolor=blue]{hyperref}
\usepackage{xspace}

\begin{document}
   \title{APODIZED PUPIL LYOT CORONAGRAPHS FOR ARBITRARY APERTURES. V.\\HYBRID SHAPED PUPIL DESIGNS for imaging Earth-like planets WITH FUTURE SPACE OBSERVATORIES}

   \author{Mamadou N'Diaye\altaffilmark{1}, R\'emi Soummer\altaffilmark{1}, Laurent Pueyo\altaffilmark{1}, Alexis Carlotti\altaffilmark{2}, Christopher C. Stark\altaffilmark{1}, Marshall D. Perrin\altaffilmark{1}}
   \altaffiltext{1}{Space Telescope Science Institute, 3700 San Martin Drive, 21218 Baltimore MD, USA}
      \altaffiltext{2}{CNRS, IPAG, F-38000 Grenoble, France}

  \begin{abstract}
We introduce a new class of solutions for Apodized Pupil Lyot Coronagraphs (APLC) with segmented aperture telescopes to remove broadband diffracted light from a star with a contrast level of $10^{10}$. These new coronagraphs provide a key advance to enabling direct imaging and spectroscopy of Earth twins with future large space missions. Building on shaped pupil (SP) apodization optimizations, our approach enables two-dimensional optimizations of the system to address any aperture features such as central obstruction, support structures or segment gaps. We illustrate the technique with a design that could reach $10^{10}$ contrast level at 34\,mas for a 12\,m segmented telescope over a 10\% bandpass centered at a wavelength $\lambda_0=$500\,nm. These designs can be optimized specifically for the presence of a resolved star, and in our example, for stellar angular size up to 1.1\,mas. This would allow probing the vicinity of Sun-like stars located beyond 4.4\,pc, therefore fully retiring this concern. If the fraction of stars with Earth-like planets is $\eta_{\Earth}=0.1$, with 18\% throughput, assuming a perfect, stable wavefront and considering photon noise only, 12.5 exo-Earth candidates could be detected around nearby stars with this design and a 12\,m space telescope during a five-year mission with two years dedicated to exo-Earth detection (one total year of exposure time and another year of overheads). Our new hybrid APLC/SP solutions represent the first numerical solution of a coronagraph based on existing mask technologies and compatible with segmented apertures, and that can provide contrast compatible with detecting and studying Earth-like planets around nearby stars. They represent an important step forward towards enabling these science goals with future large space missions. 
    \end{abstract}
 
   \keywords{Instrumentation: high angular resolution -- Techniques: high angular resolution -- Telescopes -- Methods: numerical}

   \shorttitle{APLC/Shaped Pupil Hybrid Solutions}


\section{Introduction}\label{sec:intro}

Discovering and studying nearby Earth twins with direct imaging and spectroscopy requires a telescope large enough to provide sufficient angular resolution and sensitivity to planets around a significant sample of stars \citep[e.g.][]{Stark2014}. These exoplanet goals align well with other general Astrophysics needs, and large space missions are being proposed and studied by NASA and by the community \citep[e.g. Large UV/Optical/IR (LUVOIR), Advanced Technologies Large Aperture Space Telescope (ATLAST), and High Definisiton Space Telescope (HDST)\footnote{\url{http://www.hdstvision.org/report/}}][]{Postman2012,2014NASAroadmap,Feinberg2014}. 

Because of launch systems limitations, segmented apertures are unavoidable for space telescopes larger than 8\,m typically. For starlight attenuation two types of architectures can be considered: external occulters and internal coronagraphs. External occulters, such as EXO-S\footnote{\url{http://exep.jpl.nasa.gov/stdt/Exo-S_Starshade_Probe_Class_Final_Report_150312_URS250118.pdf}} are promising solutions for exoplanet imaging and spectroscopy, with advantageous high throughput and broadband capabilities. However, for a large aperture their operations become very inefficient as their size, mass and repointing times, scale with the aperture diameter \citep{Savransky2010,Redding2014}.
In this context internal coronagraphs represent a theoretically more attractive solution for very large telescopes, but aperture discontinuities (primary mirror segmentation, central obstruction, spiders) make high-contrast imaging extraordinarily challenging with two particular major issues: starlight suppression, and high-contrast stability during observations. Given the complexity of the problem, a systemic approach is necessary to combine wavefront sensing and control for segment co-phasing and low-order aberration measurements \citep[e.g.][]{Guyon2009,Mas2012,N'Diaye2013a,Singh2014}, coronagraphy for starlight diffraction suppression \citep[e.g. see review in][]{Guyon2006}, and wavefront control for the generation and stabilization of the high-contrast dark region in the image \citep[e.g. using two deformable mirror based methods, ][]{Shaklan2006,Pueyo2007,Pueyo2009,Thomas2015arXiv}. Over the past couple of years, different research studies and laboratory testbeds were initiated to study such integrated solutions for future large space or ground-based observatories \citep{N'Diaye2013b,N'Diaye2014a,N'Diaye2015b,Galicher2014, Hicks2014,Martinez2014,Miller2015}. 



For many years, coronagraphs studies have focused on developing technologies for monolithic, off-axis telescopes. Recent progress in design and component fabrication \citep[e.g.][]{Bala2013,Delacroix2013,Otten2014,Newman2015} have allowed the community to jump over the first major hurdles, and contrasts beyond $10^9$ levels have been achieved in laboratory \citep[see review in ][]{Mawet2012}. These coronagraph designs are particularly compelling for small off-axis space telescopes that have been proposed over the past few years to observe a few hundreds of targets, directly image and characterize extrasolar planets and circumstellar disks \citep[e.g. EPIC, ACCESS, PECO, SPICES, EXO-C\footnote{\url{http://exep.jpl.nasa.gov/stdt/Exo-C_Final_Report_for_Unlimited_Release_150323.pdf}}][]{Clampin2006b,Trauger2010,Guyon2010a,Boccaletti2012,Stapelfeldt2014}. However most of these concepts are unable to perform with a large on-axis segmented telescope. 

With the development of the WFIRST-AFTA mission and its coronagraphic instrument \citep{Spergel2015,Zhao2014}, the focus has shifted to include investigations of coronagraph designs that can handle on-axis telescopes with pupil features, such as large central obstruction and support structures. Several classic coronagraphic concepts have been revisited over the past two years \citep{Baudoz2000,Rouan2000,Kuchner2002,Kasdin2003,Mennesson2003,Mawet2005,Lyon2006,Guyon2010b} and novel designs have emerged to push their performance further with arbitrary apertures \citep{Bala2008,Carlotti2011,Mawet2013,Guyon2014,Carlotti2013,Carlotti2014,Lyon2015}. In addition to the coronagraph itself, pupil remapping techniques \citep{Pueyo2013,Mazoyer2015} can be used to compensate for these aperture discontinuities with two deformable mirrors and enhanced the coronagraphic performance. Thanks to these recent progress, the selected concepts for WFIRST-AFTA are expected to reach $10^9$ contrast over a 10\% bandpass at moderate inner working angle (3\,$\lambda_0/D$) with the WFIRST-AFTA pupil \citep{Trauger2013,Guyon2014,Zimmerman2015} after image post-processing \citep[][]{Debes2015,Ygouf2015}.

The Apodized Pupil Lyot Coronagraph \citep[APLC][]{Aime2002,Soummer2003a,Soummer2005,Soummer2009,Soummer2011a} is one of the leading type of coronagraph in the current-generation ground-based instruments; it is used in the Gemini Planet Imager (GPI), Very Large Telescope (VLT) SPHERE, and Palomar P1640 \citep{Macintosh2014,Beuzit2008,Hinkley2011}. For example, the GPI APLC is designed to reach a raw contrast of $10^7$ at 0.2\,arcsec in 20\% broadband light, and in the presence of central obstructions and support structures. It uses a prolate apodization, which derives from the eigenvalue problem associated with the Lyot-style coronagraphic propagation for a given mask size. The design was optimized for the contrast as metric, by adjusting the prolate function (eigenfunction) and Lyot stop geometry to establish quasi-achromatic solutions \citep{Soummer2011a}. However, since there is only a single degree of freedom in the definition of the apodizer (the eigenvalue of the problem), the raw performance is limited to the $10^7$ to $10^8$ contrast range. 

We have recently generalized the APLC design by using a shaped pupil optimization approach \citep{Kasdin2003,Vanderbei2003a,Carlotti2011} to develop broadband solutions with raw contrast up to $10^{10}$ in one dimension \citep{Pueyo2013,N'Diaye2015a}. In addition, we have also identified solutions with this approach that increase robustness to telescope pointing errors, focus drifts and vibrations or stellar angular size. This is achieved by producing a point spread function (PSF) core smaller than the projected coronagraphic mask, allowing movement of the star PSF within the mask with virtually no impact on the coronagraph performance \citep{N'Diaye2015a}. In \citet{N'Diaye2015a}, we established the case for circular axi-symmetric obstructed pupils, using one-dimensional optimization. In that case, additional features in the pupil (e.g. support structure) deteriorate the contrast in the exoplanet search area. These effects can be partially mitigated by readjusting the Lyot stop geometry \cite{Anand2005a,Anand2005b,Soummer2009}, but this approach is limited to contrast up to $\sim10^8$.

In this paper we generalize the approach and derive new solutions for telescope apertures with arbitrary features (central obstruction, spiders, segmentation) by extending our approach from one to two-dimension (2D) pupil optimization. We present new APLCs using shaped-pupil (SP) type apodizations. These hybrid APLC/SP coronagraphs are manufacturable with current mask technologies \citep{Bala2013}. \citet{Carlotti2015} has recently proposed to combine standalone optimized shaped pupils and Lyot coronagraphs. Based on this approach, \citet{Zimmerman2015} recently proposed a design for the WFIRST-AFTA coronagraphic instrument to obtain a $10^9$ raw contrast level. In our approach, we optimize the SP apodization for the whole APLC propagation, including SP apodizer, focal plane mask, Lyot stop and final image with coronagraphic dark zone. As an application, we propose a design to produce a $10^{10}$ raw contrast with a segmented telescope aperture.  

After detailing the APLC/SP optimization methodology and our design example, we analyze the robustness of our solution to stellar angular size and derive lower space observation bounds for nearby stars with our concept. We finally estimate and discuss the yield of Earth-like planet candidates with our concept for a future mission with different telescope aperture sizes. Contrast can be assessed in many different ways, as recently recalled by \citet{Mawet2014}. In this paper and unless otherwise stated, we will refer to the raw contrast and quantify it as the averaged residual intensity over a given image area and normalized by the stellar peak intensity.

\section{Principle of Hybrid APLC/Shaped Pupil Coronagraph}\label{sec:principle}
\subsection{Formalism of Coronagraphy with Apodization}
We briefly recall the chromatic formalism of the Lyot-style coronagraph with apodization, following the notations of \citet{Aime2002} and \citet{Soummer2003a,Soummer2003b}, for a given wavelength $\lambda$ within the bandpass $\Delta\lambda$ centered at the wavelength $\lambda_0$. The vectors $\bm{r}$ and $\bm{\xi}$ (of modulus $r$ and $\xi$) denote the two-dimensional position vectors in the pupil and focal planes. 

The coronagraph setup involves four successive planes ($A$, $B$, $C$, $D$) in which $A$ defines the entrance pupil $P=P_0\,\Phi$ that combines the telescope pupil shape $P_0$ and the apodization $\Phi$, $B$ sets the location of a hard-edged focal plane mask (FPM) of diameter $m$ and transmission $1-M$ with $M(\bm{\xi})=1$ for $\xi < m/2$ and $0$ otherwise, $C$ includes the Lyot stop $L$ to filter out the diffraction due to the FPM, and $D$ represents the final image plane. The optical layout of the coronagraph is such that the complex amplitudes of the electric field in two successive planes are related by a Fourier transform. We call $\hat{f}$ the Fourier transform of the function $f$. In the absence of magnification, the electric field amplitude in the four successive planes writes as follows:
\begin{subequations}
      \renewcommand{\theequation}{\theparentequation\alph{equation}}
\begin{alignat}{2}
& \Psi_A(\bm{r}, \lambda)=P(\bm{r}) \label{subeq:Psi_A}\\
& \Psi_B(\bm{\xi}, \lambda)= \frac{\lambda_0}{\lambda}\widehat{\Psi}_A \left (\frac{\lambda_0}{\lambda}\bm{\xi}, \lambda \right ) \left (1- M(\bm{\xi})\right ) \label{subeq:Psi_B}\\
& \Psi_C(\bm{r}, \lambda)= \left (\Psi_A(\bm{r}, \lambda) - \Psi_A(\bm{r}, \lambda)* \frac{\lambda_0}{\lambda}\widehat{M}\left (\frac{\lambda_0}{\lambda}\bm{r}, \lambda \right ) \right )L(\bm{r}) \label{subeq:Psi_C}\\
& \Psi_D(\bm{\xi}, \lambda)= \left (\frac{\lambda_0}{\lambda}\right )^2 \left (\widehat{\Psi}_A\left (\frac{\lambda_0}{\lambda}\bm{\xi}, \lambda \right ) (1- M(\bm{\xi})) \right )*\widehat{L}\left (\frac{\lambda_0}{\lambda}\bm{\xi}, \lambda \right ) \label{subeq:Psi_D}
\end{alignat}
\end{subequations}
where $*$ symbolizes the convolution operator.
In the absence of FPM, the electric field amplitude in the final image plane is simply given by:
\begin{equation}
\Psi_D^0(\bm{\xi}, \lambda)= \left (\frac{\lambda_0}{\lambda}\right )^2 \widehat{\Psi}_A\left (\frac{\lambda_0}{\lambda}\bm{\xi}, \lambda \right )*\widehat{L}\left (\frac{\lambda_0}{\lambda}\bm{\xi}, \lambda \right ) 
\label{subeq:Psi_0}
\end{equation}
Assuming a flat stellar spectrum, the coronagraphic image intensity over the bandpass $\Delta\lambda$ can be deduced from Equation (\ref{subeq:Psi_D}) by means of an operator $\mathcal{I}_{\Delta\lambda}$ as
\begin{equation}
\mathcal{I}_{\Delta\lambda}[\Phi(\bm{r})](\bm{\xi})=\frac{1}{\Delta\lambda}\int_{\Lambda} \left |\Psi_D(\bm{\xi}, \lambda)\right |^2\,d\lambda\,,
\label{eq:I_Dbroad}
\end{equation}
where $\Lambda$ defines a set of wavelengths $\lambda$ such that $|\lambda-\lambda_0|< \Delta\lambda/2$.

The analysis of the re-imaged pupil plane offers insights on the coronagraph behavior. According to Equation (\ref{subeq:Psi_C}), perfect starlight suppression is obtained if the direct wave (entrance pupil amplitude) and the wave diffracted by the FPM (convolution integral) matches within the Lyot stop. Different approaches have been considered to obtain a perfect subtraction between these two waves. A first approach consists of modifying the geometry of the mask to obtain perfect subtraction within a reduced area of the pupil. Such strategy corresponds to the band-limited mask coronagraph \citep{Kuchner2002}. We follow a second approach seeking for a modification of the entrance pupil amplitude by means of the apodization $\Phi$ to match the two waves within the re-imaged pupil \citep{Aime2002,Soummer2003a}.

The existence of a perfect extinction solution for the Roddier phase mask coronagraph \citep{Roddier1997} in monochromatic light was numerically derived by \citet{Guyon2000} and analytically showed by \citet{Aime2002} and \citet{Soummer2003a}. For the APLC, there is no solution for complete starlight suppression but \citet{Aime2002} and \citet{Soummer2003a} analytically show that prolate apodization solutions nearly achieve complete starlight removal with circular and rectangular unobstructed apertures. These solutions have been generalized to the obstructed circular aperture \citep{Soummer2005} and the arbitrary non circular aperture in the context of Extremely Large Telescopes \citep[ELTs,][]{Soummer2009,Martinez2007}. These solutions for opaque masks have also been extended to the complex amplitude FPMs for small IWA coronagraphy \citep{Guyon2010b,Guyon2014}. 

In the case of opaque FPM, \citet{Soummer2011a} studied the chromatic properties of APLC using prolate apodizers and found the existence of quasi-achromatic solutions over large bandpasses. They derive a continuous family of prolate apodizations corresponding to the telescope geometry and select the optimal set of apodization and Lyot stop geometry using contrast in the final image plane within the controllable adaptive optics (AO) area as an optimization metric to obtain the broadband APLC designs that are currently implemented in P1640 and GPI \citep{Hinkley2011,Macintosh2014}. 

The raw performance of the current implementations are however limited by the available degrees of freedom (apodization and Lyot stop geometry) in the optimization approach. We recently found new functioning modes for the APLC by proposing an alternative approach based on the shaped-pupil like optimization to design Lyot-style coronagraph with improved performance in terms of contrast and inner working angle \citep{Pueyo2013,N'Diaye2015a}. In \citet{N'Diaye2015a}, we propose broadband solutions for centrally obstructed circular apertures, using 1D pupil optimization. We here show the existence of solutions for arbitrary apertures with pupil features such as segmentation and spiders by exploiting our approach to 2D pupil.

\subsection{Numerical Optimization for the Apodizer}
We define a numerical optimization problem with Lyot-style coronagraph that includes the aperture, FPM, and Lyot Stop geometries, and constraints on the coronagraphic intensity in a given focal plane region (dark zone) and we look for apodization solutions for the APLC. Among these solutions, we seek for the apodization with the best throughput in an attempt to maximize the number of photons from companions around the observed star during observing time. Following the classical shaped-pupil type optimization problems \citep{Vanderbei2003a,Kasdin2003,Carlotti2011}, the optimal pupil apodization for APLC can be found by solving the following optimization problem:
\begin{subequations}
      \renewcommand{\theequation}{\theparentequation\alph{equation}}
\begin{alignat}{2}
& \max \left [\int_{P_0} \Phi(\bm{r})d\bm{r}\right ] \quad \label{subeq:maxintP} \text{under the contraints:}\\
& \left |\Psi_D(\bm{\xi}, \lambda)\right |^2 < 10^{-C} \left |\Psi_D^0(\bm{0}, \lambda_0)\right |^2 \quad \text{for} \quad
\begin{dcases}
\rho_i < \xi < \rho_o\\
|\lambda-\lambda_0| < \Delta\lambda/2 \label{subeq:C} 
\end{dcases}\\
& 0 \leq \Phi(\bm{r}) \leq 1 \label{subeq:maxP},
\end{alignat}
\end{subequations}
where the significance of each line is further explained below:

(\ref{subeq:maxintP}) The apodizer throughput is a quadratic function of the field amplitude. Assuming  positive, real-valued apodization amplitude (see Eq. (\ref{subeq:maxP})), we use this linear function of the field amplitude as a proxy to maximize the apodizer throughput over the area defined by the initial aperture shape $P_0$. 

(\ref{subeq:C}) The coronagraphic image intensity is constrained at all the wavelengths over the bandpass to achieve a contrast $10^C$ over the area $\mathcal{D}$ in broadband light. This region is defined between its inner and outer edges, respectively $\rho_i$ and $\rho_o$. 

(\ref{subeq:maxP}) The transmission of the amplitude apodization is real, positive and normalized.

\citet{N'Diaye2015a} set constraints on the apodizer derivative to obtain continuous solutions with limited oscillations within the pupil apodization and avoid bang-bang solutions. 
The problem was written in one dimension with linear constraints, allowing resolution with a linear programming solver \citep[][]{Vanderbei2009}

The 2D problem does not include constraints on the apodizer derivative, thus enabling binary apodization solutions using shaped-pupil manufacturing technologies. While the 2D optimization problem can also be expressed linearly as the 1-Dimensional problem, we find that this problem in AMPL language \citep{Fourer1990} with the solver Gurobi \citep{gurobi} is handled slightly more effectively using the image intensity instead of the field as showed in Eq. (\ref{subeq:C}) from described above. This somewhat counter-intuitive behavior is likely to be solely due to technical implementation of these various solvers.

The computation of the coronagraphic electric field for the APLC requires three successive Fourier transforms (FT) between the electric field amplitude in the four successive planes of the optical layout. We adopt the semi-analytical approach from \citet{Soummer2007} for fast Lyot-style coronagraph computation with simultaneous highly sampled pupils and focal plane masks. The semi-analytical and FFTs approaches are strictly equivalent in the absence of field of view limitations in the FPM plane. Further error budget studies should provide the acceptable size of the mirror for a reflective FPM to obtain a compatible  $10^{10}$ contrast level. We also express the Fourier transforms (FTs) involved in the coronagraphic electric field calculation using discrete Riemann sum in Cartesian coordinates. Following \citet{Vanderbei2012}, these 2D FTs are performed in two steps with a separation of the variables to reduce the problem complexity. \citet{Carlotti2011} successfully applied this strategy to design shaped pupils in a two-plane configuration that simply involves a single FT for the coronagraphic electric field computation. 

We here seek solutions based on an apodizer and focal plane mask with real amplitude transmission (no phase terms). Further solutions with complex amplitude apodizer and focal plane masks could also be investigated based on the optimization problem set above without loss of generality, but this study is out of the scope of this paper.

\section{Application to a Large Segmented Aperture}\label{sec:application}
\subsection{Solution}
To illustrate these new APLC/SP hybrid solutions, we present a design optimization for a 37-hexagonal-segment telescope aperture with central obstruction and spiders, such as one of the envisioned design for the ATLAST mission concept \citep{Postman2012} or the geometry corresponding to existing ground-based telescopes W.M. Keck, or Gran Telescopio Canarias. In the following, all the pupil geometry parameters (obstruction, spiders, gaps) are expressed in aperture diameter ratio.

\begin{figure}[!ht]
\centering
\resizebox{\hsize}{!}{
\includegraphics{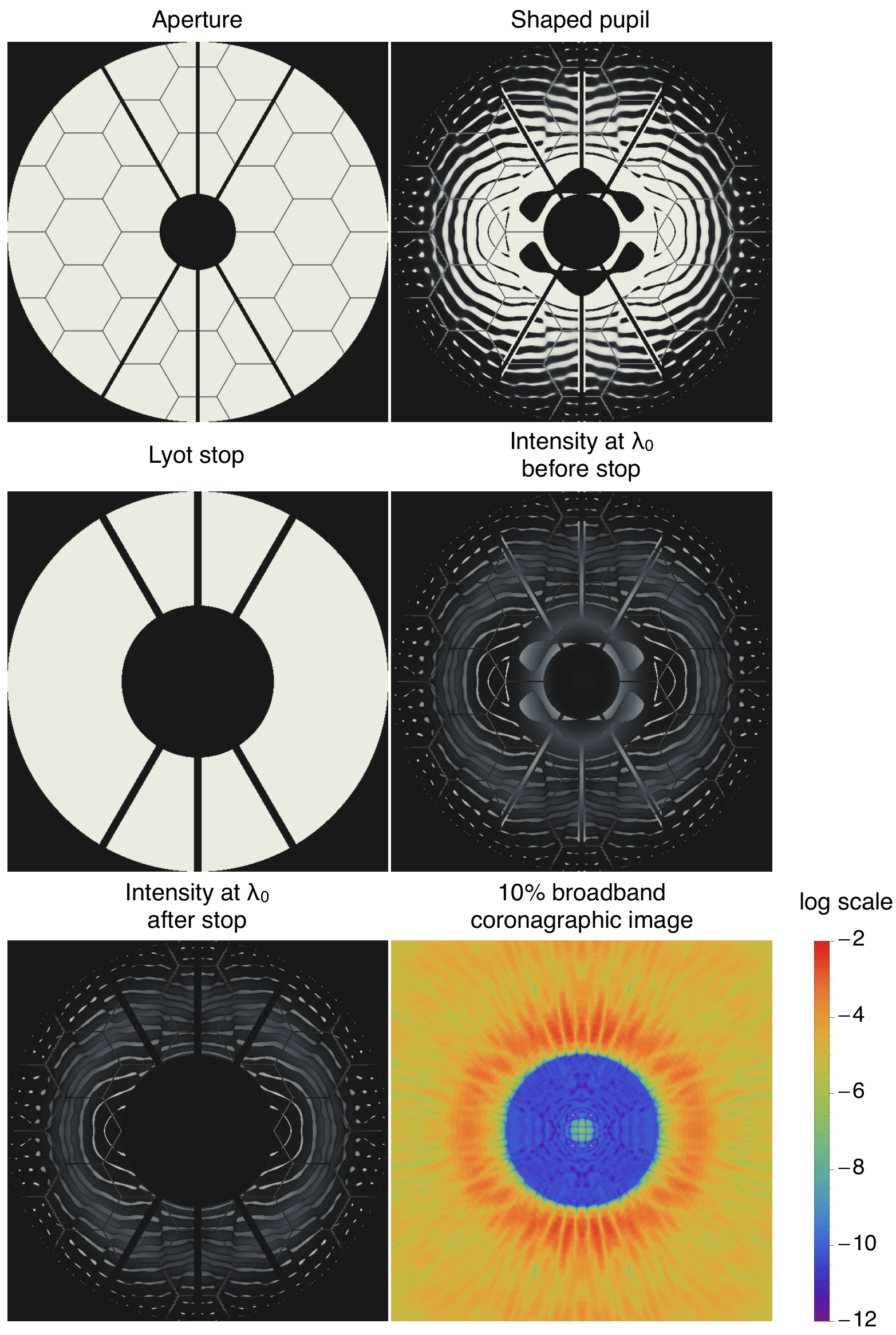}
}
\caption{Top left: telescope aperture. Top right: Shaped pupil apodization for the APLC with the previous aperture. Middle Left: Lyot stop shape. Middle right and bottom left: Residual intensity in the relayed pupil plane before and after Lyot stop application. Bottom right: Broadband coronagraphic image of the star with the APLC/shaped pupil hybrid design using the shaped pupil and Lyot stop showed above, and the FPM of radius $r=4\,\lambda_0/D$. A $10^{10}$ contrast region with 3.5\,$\lambda/D$ inner and 20\,$\lambda_0/D$ outer edges is achieved with our solution over a 10\% spectral bandpass, paving the way for the direct imaging of Earth twins with future large missions.}
\label{fig:atlast_corono_design}
\end{figure}

We assume an inscribed circular aperture over the 37 segments, with 20\% central obstruction, 1\% spiders and 0.2\% segments gaps, see Figure \ref{fig:atlast_corono_design} top left. For the coronagraph design, we consider an FPM of radius $m/2=4\,\lambda_0/D$. The contrast target is set to $10^{10}$ in a focal plane region $\mathcal{D}$ delimited by inner and outer edge radii of $\rho_i=$3.5 and $\rho_o=$20.0\,$\lambda_0/D$. Following the approach developed in \citet{N'Diaye2015a}, we set the dark zone inner edge with a smaller radius than the FPM. Such resulting designs do not allow observations of companions at separations smaller than the projected FPM angular size but the coronagraphic solution will present a reduced sensitivity to low-order aberrations, such as telescope jitter and thermal or mechanical focus drift, as well as a reduced sensitivity to the stellar diameter. 

We finally consider a circular Lyot stop with the same outer size as the entrance aperture, with 40\% central obstruction and 2\% spiders, corresponding to an oversizing factor of 2 with respect to the entrance pupil characteristics, but without reticula for the segment gaps (no obstruction for these aperture discontinuities), see Figure \ref{fig:atlast_corono_design} middle left. To achieve a broadband APLC design, we run an optimization with constraints at three wavelengths that are centered around $\lambda_0$ and equally spatially sampled one to another over 10\% bandwidth. Note that while only three wavelengths are used in this case for the optimizer, the broadband profiles illustrated in the figures include 11 simulated wavelengths. The parameters are summarized in Table \ref{table:params}.

With these parameters, we solve our optimization problem and find an APLC design solution with the shaped-pupil apodization displayed in Figure \ref{fig:atlast_corono_design} top right, leading to a 28\% total coronagraph throughput. Our solution produces the relayed pupil image that is showed at $\lambda_0$ before and after the Lyot stop application in the middle right and bottom left frames. As expected, most of the star diffracted light is found at the aperture discontinuities \citep{Anand2005a,Anand2005b} and constrained within the area blocked by the Lyot stop geometry, allowing an efficient starlight rejection by the Lyot Stop.
It is interesting to note that the introduction of a Lyot stop reticula \citep{Anand2005a,Anand2005b} in this specific design is counterproductive in terms of starlight suppression: it results in additional diffracted light within the final image plane and a deterioration of the contrast in the dark region. An APLC solution using a Lyot stop with reticula can be designed with our approach if the geometry of the diaphragm is included in the optimization process.

\begin{figure}[!ht]
\centering
\resizebox{\hsize}{!}{
\includegraphics{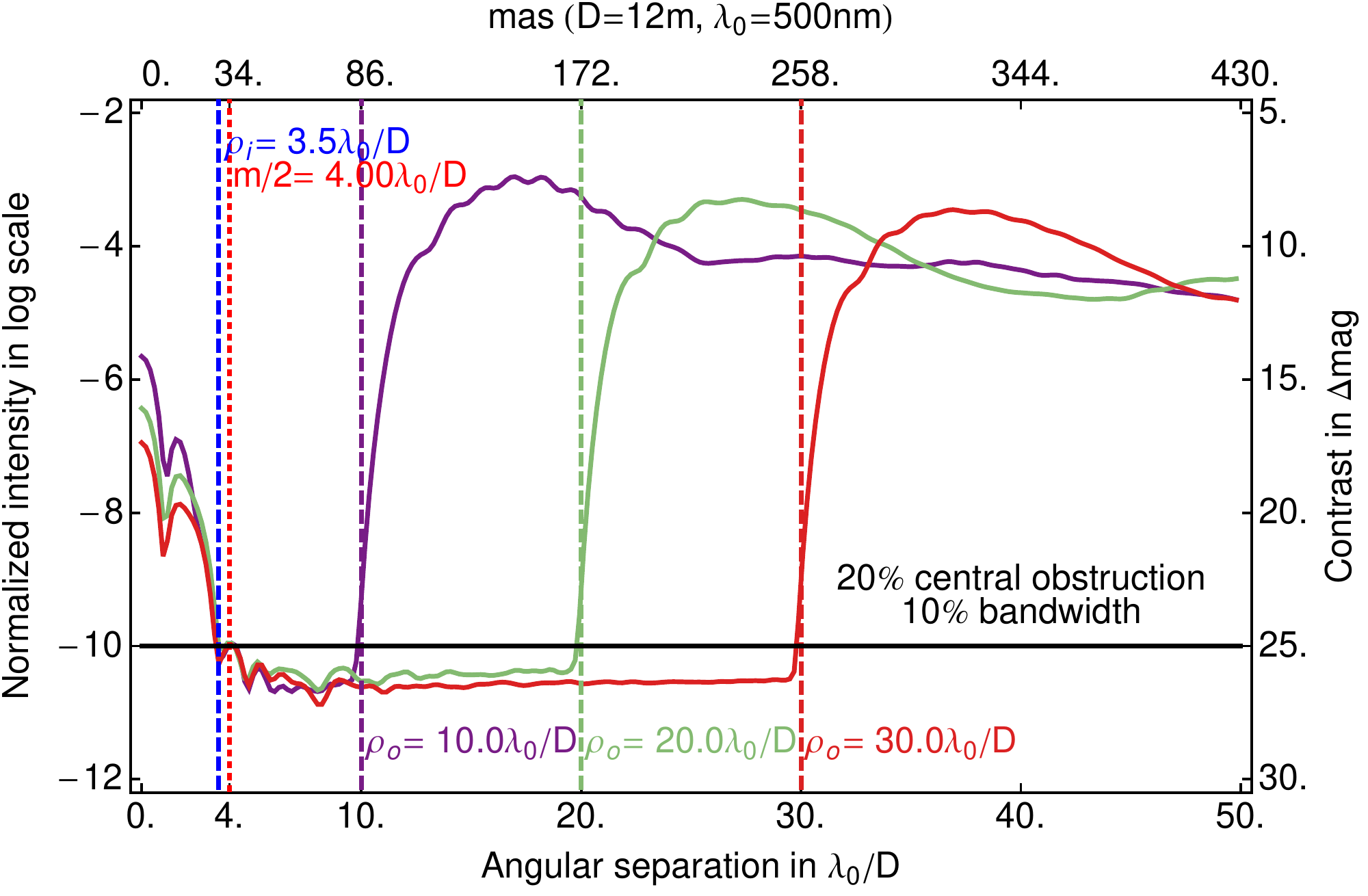}
}
\caption{Azimuth averaged intensity profile of the coronagraphic image produced by our APLC/SP hybrid solutions in 10\% broadband light with the components showed in Figure \ref{fig:atlast_corono_design}, for three different dark zone outer edges. The vertical dashed lines represent the bounds of the search area $\mathcal{D}$ ($\rho_i=3.5\,\lambda_0/D$ [blue] and $\rho_o=$10.0, 20.0, and 30.0\,$\lambda_0/D$ [purple, green, red]). The red dot line delimits the FPM radius, set to $m/2=$4\,$\lambda_0/D$. The averaged contrast over the spectral band in the dark region is below $10^{-10}$ (black horizontal line). For each dark zone outer bound, the coronagraph has here been designed to produce an image core smaller than the projected FPM size, allowing for both enlargement and expansion of the PSF without impact on the contrast, making this design virtually insensitive to low-order aberration, as explained in \citet{N'Diaye2015a}.}
\label{fig:atlast_corono_psf}
\end{figure}

The broadband coronagraphic image and its corresponding averaged intensity profile are displayed in Figures \ref{fig:atlast_corono_design} bottom right and \ref{fig:atlast_corono_psf} for designs with three different dark zone outer bound, illustrating the ability of our approach to design APLCs that could produce broadband PSFs with $10^{10}$ dark region in the presence of an arbitrary aperture. 

\begin{table}[!ht]
\caption{Parameters for the design showed in Figure \ref{fig:atlast_corono_design}.}
\centering
\begin{tabular}{c c}
\hline\hline
parameters & value\\
\hline
Contrast C & $10^{10}$\\
Dark zone inner edge radius $\rho_i$ & 3.5\,$\lambda_0/D$\\
Dark zone outer edge radius $\rho_o$ & 20.0\,$\lambda_0/D$\\
focal plane mask radius $m/2$ & 4.0\,$\lambda_0/D$\\\hline
Aperture & 20\% central obstruction\\
& 1\% aperure size spiders\\
& 0.2\% aperture size gaps\\\hline
Lyot stop & 40\% central obstruction\\
& 2\% aperutre size spiders\\
& no segmentation\\\hline
Broadband optimization & 3 wavelengths within 10\% band\\
Broadband PSF profiles calculations & 11 wavelengths within 10\% band \\
\hline
\end{tabular}\\
\label{table:params}
\end{table}

\subsection{Apodizer Manufacturing Aspects}
Because of numerical limitations for the number of points in the pupil, our APLC/SP optimization produces an apodization which values are mostly black or white (37\% of the apodization points have a 0 or 100\% amplitude transmission level to a 1\% precision) but that also contains gray points. We investigated the conversion of these apodization solutions into binary apodizers that can be manufactured using the current state-of-the-art technologies developed for shaped pupil apodizers \citep{Bala2013}.

Our grey designs are obtained with a 600-point diameter. A pupil sampling increase or a full exploration of the coronagraph parameter space (FPM size, Lyot stop geometry, dark zone dimensions) could allow us to find fully black and white designs but proves computationally challenging. Instead, we investigated new versions of our original grey design, using different methods to convert our grey pixels into black and white, such as simple rounding of the values or a sub-pixelization of the gray design pixels followed by application of the error diffusion algorithm \citep[EDA, ][]{Floyd1976}, a process that has been used for the manufacturing of grey apodizers for GPI, SPHERE and P1640 to reach $10^6-10^7$ raw contrast levels \citep{Dorrer2007,Martinez2009a,Anand2009}. At $10^{10}$ raw contrast levels, the same process is expected to be efficient if the performance of the binary design is validated numerically and then experimentally and as long as the pixel size of the apodizer prototype remains significantly larger than the wavelength of observation.

\begin{figure}[!ht]
\centering
\resizebox{\hsize}{!}{
\includegraphics{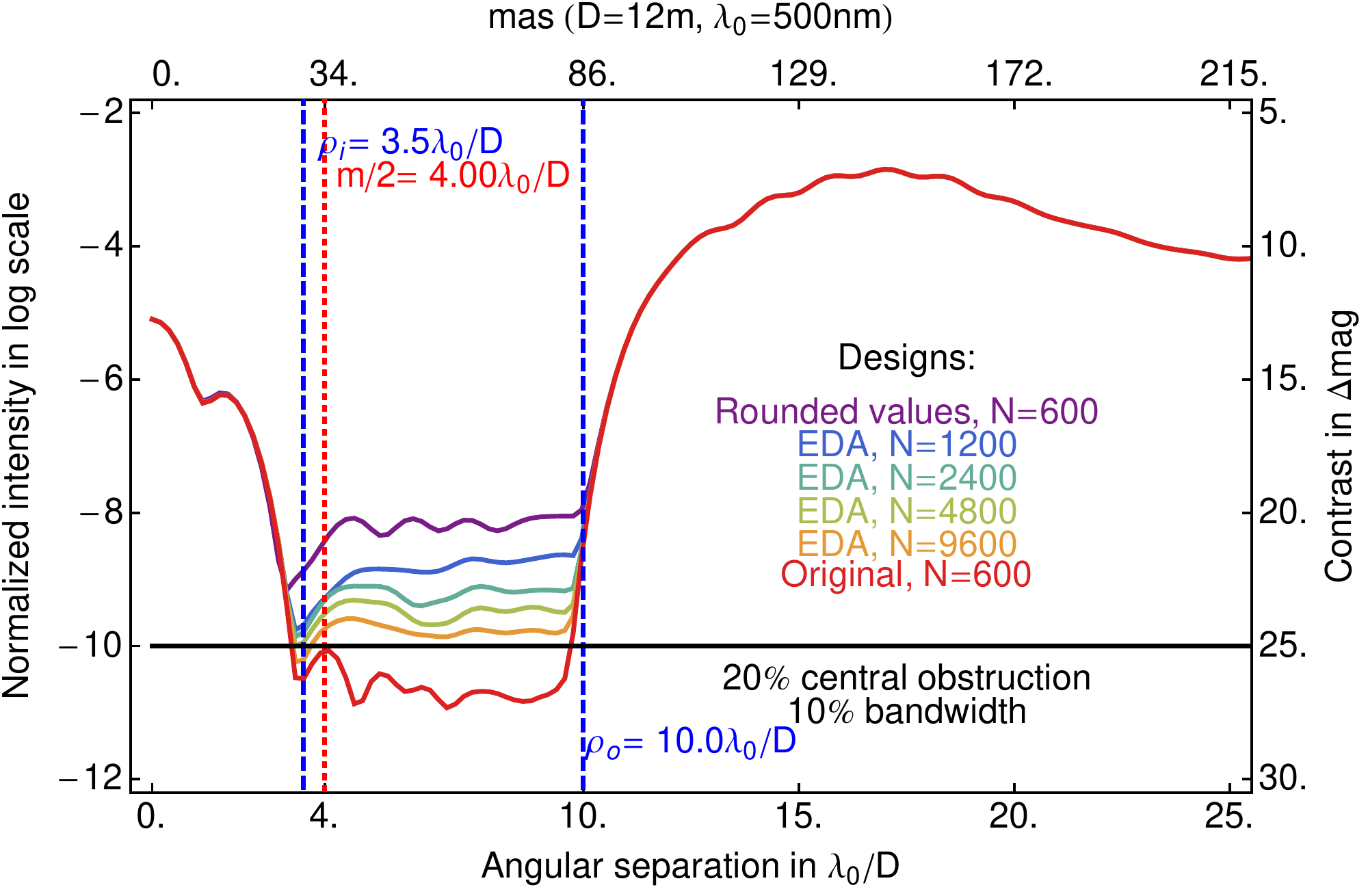}
}
\caption{Azimuth averaged intensity profile of the coronagraphic image reached by an APLC solution in 10\% broadband light and the parameters shown in Table \ref{table:params}, except for the Lyot stop which presents a 36\% instead of 40\% obstruction, providing a quasi-binary shaped pupil. Profiles are represented for the design with the original grey version, rounded values version, and the binarized versions using error diffusion algorithm (EDA, see e.g. \citet{Dorrer2007}) with different lateral size $N$ obtained by sub-pixelization of the original design gray pixels. The dashed blue vertical lines delimit the high-contrast search area $\mathcal{D}$ ($\rho_i=3.5\,\lambda_0/D$ and $\rho_o=$10.0\,$\lambda_0/D$). The red dot line delimits the FPM radius, set to $m/2=4\,\lambda_0/D$. The averaged contrast over the spectral band in the dark region is below $10^{-10}$ (black horizontal line). The $10^{10}$ contrast performance of the original design is almost recovered with a EDA version using N=9600. Relating on current black silicon technologies to manufacture Shaped pupil mask for WFIRST-AFTA coronagraph \citep{Bala2013}, we translate these values into physical units. Assuming a 10\,$\mu$m size for a pixel, the apodizer of our design can currently be fabricated with a 96\,mm diameter prototype to work in visible light.}
\label{fig:atlast_design_binarity}
\end{figure}

Figure \ref{fig:atlast_design_binarity} displays the contrast curves for our coronagraph design with a dark zone outer edge $\rho_1$=10\,$\lambda_0/D$ using the original apodizer from the optimizer and its binarized versions from numerical simulations. A contrast loss of two orders of magnitude is observed by simply rounding the pixel values. The contrast loss can be recovered by using the EDA method and sufficient oversampling. With a factor of 16 for the sub-pixelization, we almost recover the initial $10^{10}$ contrast level reached with our optimized design. With this 9600-pixel diameter apodizer, we translate these values into physical units for manufacturing considerations. We assume the use of black silicon technologies with the current state-of-the-art for the fabrication of a reflective apodizer prototype \citep{Bala2013}. Assuming a 10\,$\mu$m pixel size, an apodizer prototype with 96\,mm diameter can be manufactured fulfilling the specifications that enable the starlight suppression down to $10^{-10}$ intensity level in visible light with a telescope segmented aperture. This result shows the existence of suitable, broadband diffraction suppression solutions with already manufacturable components (shaped-pupil apodization, FPM, Lyot stop) for the direct imaging and spectral analysis of Earth twins with large segmented apertures. Further coronagraph designs using binary apodization with a smaller sampling are currently under investigation to reduce the pupil dimensions in the context of a coronagraph instrument.

\subsection{Off-axis throughput}
Beyond starlight removal, estimating the amount of photons from the nearby objects through the coronagraph is important to determine the observation time required for accurate photometry, astrometry, and spectral analysis of the star companions. We consider the Airy throughput as a metric to determine the fraction of transmitted light from off-axis companions through our coronagraphic system. The Airy throughput is defined as the ratio of the total energy inside the core of the off-axis coronagraphic PSF to the total energy inside the core of the telescope PSF. Since our design makes use of a pupil apodization, the image of an off-axis companion is an apodized PSF, which sharpness is defined by the shaped pupil. Throughputs are displayed for our coronagraph designs with three different PSF dark zones in Table \ref{table:throughput}. For the Airy throughput, we adopt a 0.75\,$\lambda_0/D$ core radius for the photometric aperture to account for the modified light distribution in the PSF in the presence of the shaped-pupil apodization. This ideal value has been determined after calculating the exo-Earth yield, presented in the next section, as a function of the photometric size aperture to maximize the planet's light signal compared with different sources of noise backgrounds (leaked starlight, exozodi, local zodi, etc.). In the background-limited regime, for an Airy pattern PSF, the ideal photometric aperture radius is $\sim$ 0.7\,$\lambda_0$/D, which contains $\sim$ 70\% of the planet's PSF.

Our throughputs have been estimated with a sampling of 5 pixels per resolution element ($\lambda_0/D$). For the design showed in Figure \ref{fig:atlast_corono_design}, the Airy throughput within the dark region and with respect to the segmented aperture is estimated with a maximum of 18.4\% and with about half of it at an angular separation of 4.2\,$\lambda_0/D$, giving the IWA of our design. Table \ref{table:throughput} summarizes the throughputs for our coronagraph designs, showing its evolution as a function of different separations.

Further improvements in contrast and IWA are expected with a full exploration of the parameter space for this coronagraph (FPM, Lyot stop geometry, bandwidth), following \citet{N'Diaye2015a}, to identify the solution with the best throughput for a given aperture geometry.

\begin{table}[!ht]
\caption{Throughput estimates for our coronagraph designs in Figure \ref{fig:atlast_corono_psf} with different dark zone outer bounds. Airy throughputs are estimated with a 1.5\,$\lambda_0/D$ size photometric aperture , using 5 pixels per $\lambda_0/D$.}
\centering
\begin{tabular}{c c c c c c c c}
\hline\hline
\multicolumn{2}{c}{Values wrt to aperture geometry} & \multicolumn{3}{c}{Clear} & \multicolumn{3}{c}{Segmented}\\ 
\multicolumn{2}{c}{$\rho_o$ in $\lambda_0/D$} & 10.0 & 20.0 & 30.0 & 10.0 & 20.0 & 30.0\\
\hline
\multirow{4}{*}{Airy throughput in \% at} & 4.0\,$\lambda_0/D$ & 3.1 & 2.9 & 2.7 & 5.8 & 5.4 & 5.1\\
 & 4.2\,$\lambda_0/D$ & 5.1 & 4.7 & 4.5 & 9.5 & 8.9 & 8.4\\
 & 5.0\,$\lambda_0/D$ & 8.6 & 8.1 & 7.7 & 16.1 & 15.2 & 14.4\\
 & 6.0\,$\lambda_0/D$ & 10.0 & 9.3 & 8.8 & 18.7 & 17.5 & 16.5\\
 \multicolumn{2}{c}{Max. Airy Throughput in \%} & 10.5 & 9.8 & 9.3 & 19.8 & 18.4 & 17.4\\
\hline
\end{tabular}
\label{table:throughput}
\end{table}

\subsection{Stellar Angular Size Sensitivity Analysis}
Since on-axis resolved stars is identified as one of the major concerns for high-contrast imaging of extrasolar planets with future large telescopes \citep{Guyon2006}, we analyze the impact of the stellar angular size on the performance of our coronagraphs.

An on-axis resolved star can be seen as a combination of off-axis point sources at all the angular positions ranging up to the stellar angular radius. This is equivalent to consider a sum of tip-tilt point sources weighted by $2\,\pi\,\xi$ where $\xi$ here corresponds to the offset position. We can devise the coronagraph contrast sensitivity at 5\,$\lambda_0$/D by averaging the solid-line profiles for the coronagraphic images of tip-tilt sources with weights corresponding to the offset positions, as previously done in \citet{N'Diaye2015a}.

\begin{figure}[!ht]
\centering
\resizebox{\hsize}{!}{
\includegraphics{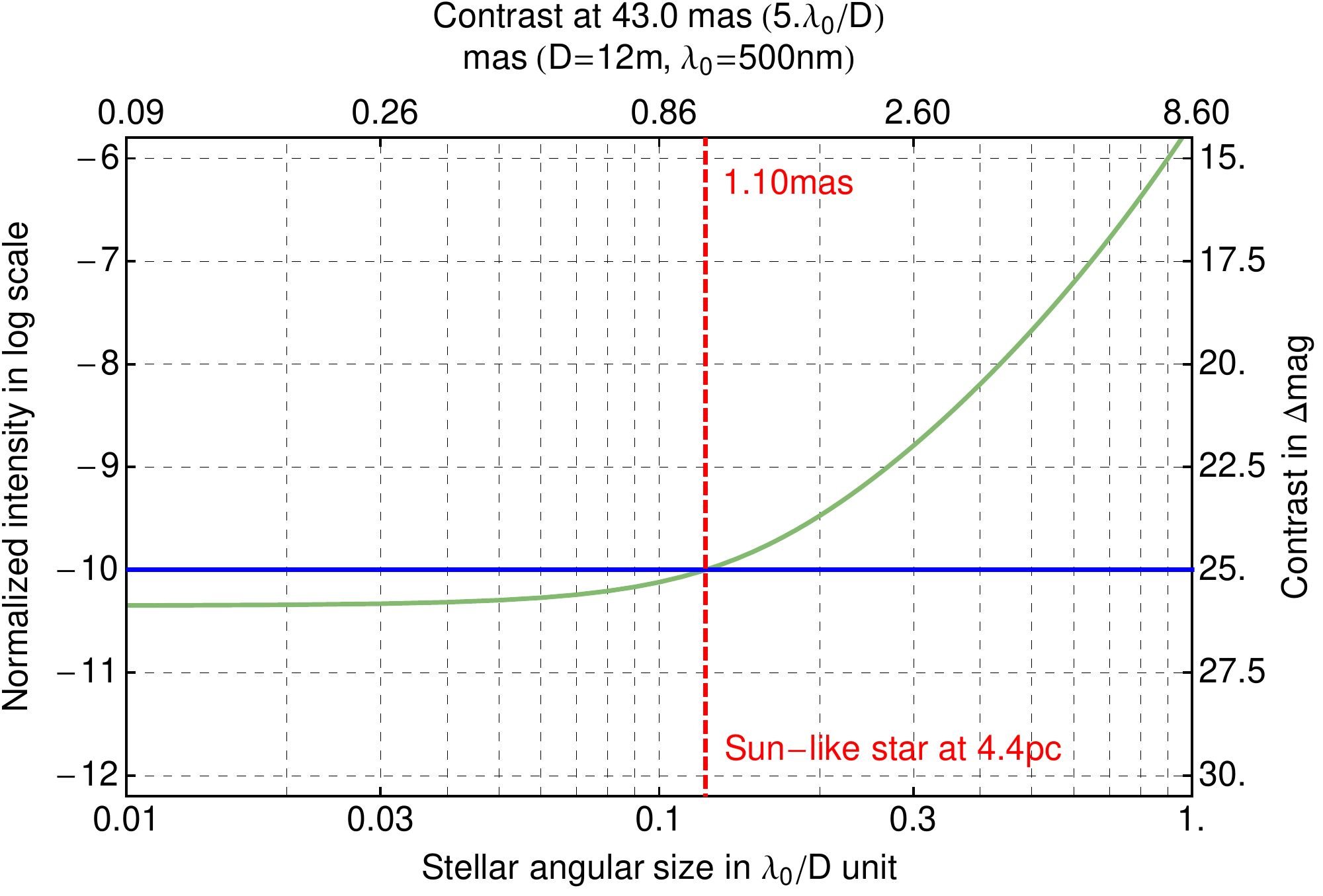}
}
\caption{Averaged intensity of the broadband coronagraphic image at a 5\,$\lambda_0/D$ angular separation from the optical axis as a function of the stellar angular size. The curve is obtained for our design with $\rho_1=$20\,$\lambda_0/D$. Blue solid line denotes the $10^{-10}$ intensity level. Our design presents a plateau and intensity levels below $10^{-10}$ for stellar angular size up to 0.12 $\lambda_0/D$, underlining the contrast performance stability of our coronagraph design. Assuming a 12\,m telescope at 500\,nm, our coronagraph is robust to stellar angular sizes up to 1.1\,mas, allowing observations of planets around Sun-like star located beyond 4.4\,pc.}
\label{fig:atlast_star_size}
\end{figure}

Figure \ref{fig:atlast_star_size} shows the contrast performance at 5\,$\lambda_0$/D of the 20\,$\lambda_0/D$-OWA design as a function of the angular size of the observed objects. Our designs are robust enough to stellar angular size to allow observation of planets around Sun-like stars located beyond 4.4\,pc.

\begin{figure}[!ht]
\centering
\resizebox{\hsize}{!}{
\includegraphics{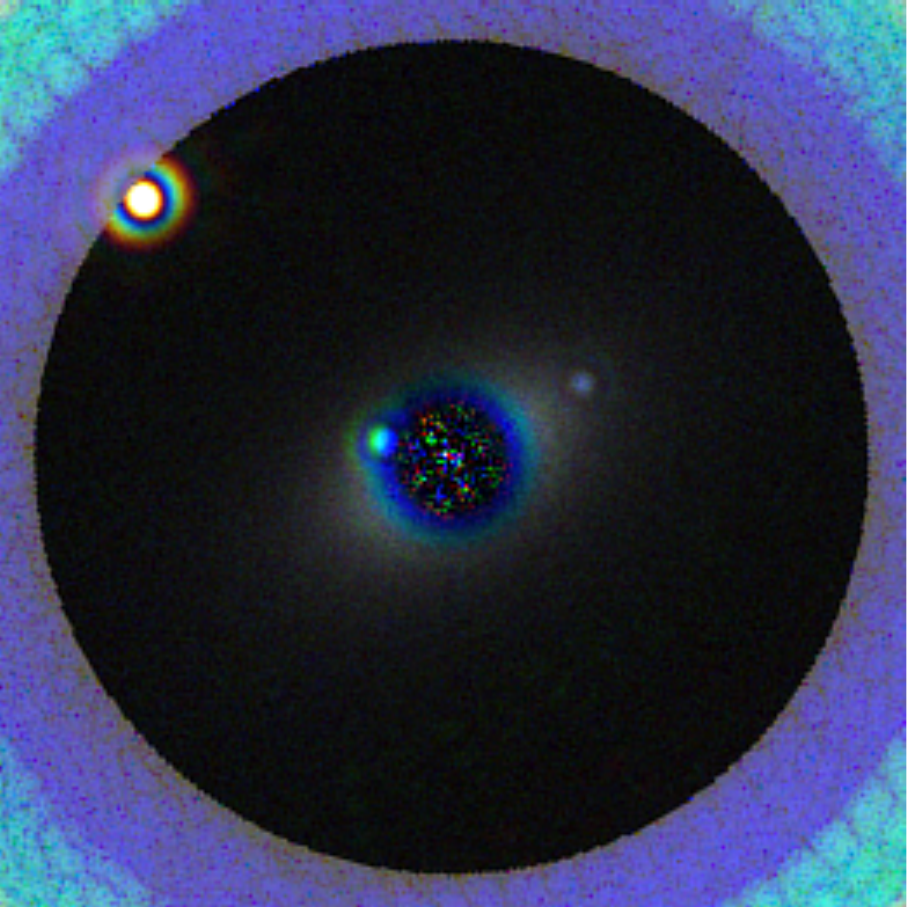}
\includegraphics{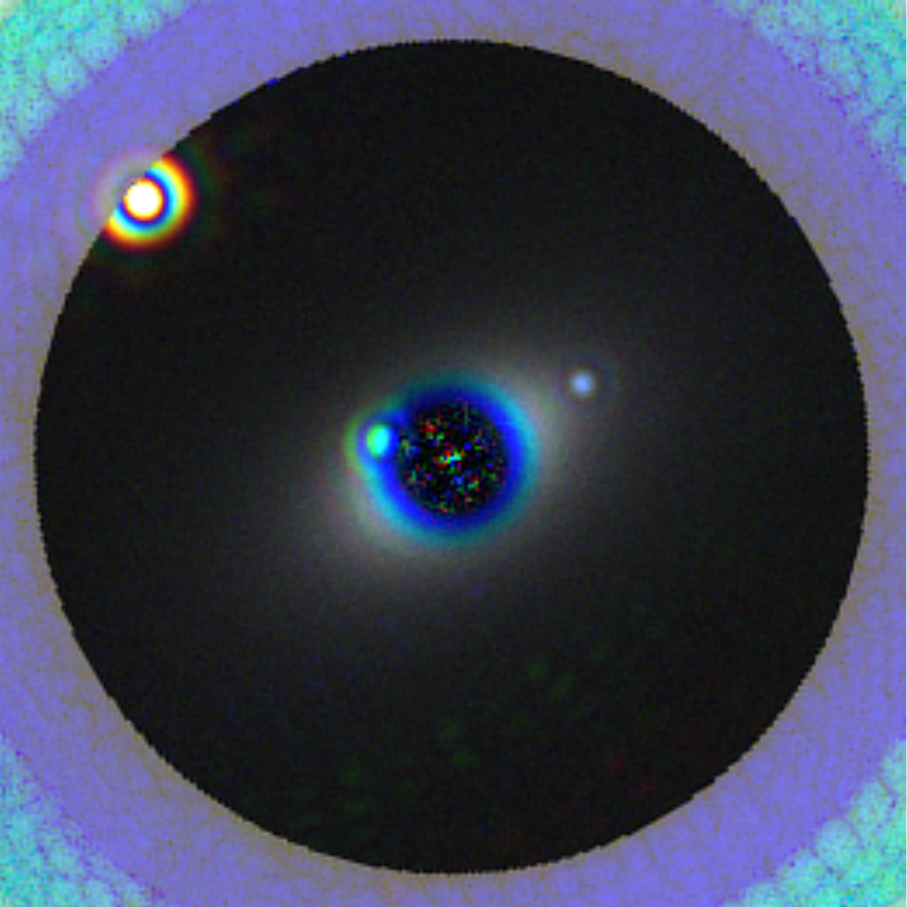}
}
\caption{Simulated multicolor pictures of a solar system twin at 13.5 pc with a 12 m telescope and our coronagraph design with an OWA $\rho_1=$30\,$\lambda_0$/D at different exposure times (from left to right: 8h, 40h). The solar system was modeled with the Haystacks project (Courtesy A. Roberge et al.), including spectroscopic and spatial information for all the components within the exoplanetary system. Each picture is a composite of three images on three channels around 400, 500 and 600 nm, each with 10\% bandpass and a coronagraph optimized for this band. In this simulation, we assume a perfect wavefront, no wavefront drifts between the target “solar system” star and calibrator star for the image processing, and only photon noise. Earth is at 2 o'clock and is indeed blue, Venus is at 9 o'clock, Zodiacal light is elongated along the 2-8 o'clock direction, Jupiter is at 10 o'clock in the red channel (600 nm): it is outside of the dark hole at shorter wavelengths. In that channel, most of Venus is hidden by the coronagraph. Linear and logarithmic scale representations are used inside and outside the high-contrast region for each coronagraph.}
\label{fig:multicolor_picture}
\end{figure}

With our hybrid concept and assuming a perfect and stable wavefront, a solar system twin with unresolved companions at 13.5\,pc could be observed with a 12\,m telescope in 8\,h and characterized with less than two-day exposure time under photon-noise limited regime, see Figure \ref{fig:multicolor_picture}. In this numerical simulation, we did not account for the motion of the planet and the required post-processing methods to increase the signal-to-noise ratio of planets in such long period observations. Promising strategies for such effect have recently been proposed to efficiently combine the different frames of the sequence of observation for the planet with orbital motion during the observations \citep{Males2015}. Our design is fully functioning for detection and characterization of Earth twins around nearby stars.

\section{Exo-Earth Yield Estimate}\label{sec:AYO}
The number of observable Earth-sized planets in the habitable zone of their parent star (exo-Earth candidates) represents a key metric for future exoplanet imaging missions. Several studies have estimated the exo-Earth candidate yield with future coronagraphic missions \citep[e.g.][]{Brown2005,Brown2010}. We estimate the exo-Earth candidate yield for a future exoplanet direct imaging mission with our coronagraph design using the methods developed by \citet{Stark2014} and \citet{Stark2015}.  These methods maximize the yield of Earth twins by simultaneously optimizing many aspects of the observation plan, including the exposure times of all observations, the number of visits to each star, and the delay time between revisits.

For our design reference missions analysis, we assume a five-year mission using our coronagraph design with two years dedicated to exo-Earth detection (one total year of exposure time and another year of overheads). We assume the dimmest detectable planet is 27.5\,mag fainter than its host star, one order of magnitude beyond the raw contrast. We note that although our adopted noise floor is very optimistic, noise floors beyond $\sim$26\,mag do not greatly impact the yield; adopting a more conservative noise floor of 26\,mag fainter than the host star will not change our results significantly \citep{Stark2015}. We assume an observation time only devoted to detection in V band (no spectral characterization) with signal-to-noise ratio SNR$\geq$7, and noiseless detectors (photon noise limited observations).

We use several astrophysical assumptions to estimate our Earth-analog candidate yield, see \citet{Stark2015} for a detailed description of astrophysical assumptions made here. We assume Earth twins with parameters as described in \citet{Stark2014}. We adopt the optimistic habitable zone definition from \citet{Kopparapu2013} for Sun-like stars (OKHZ), ranging from 0.75 - 1.77\,AU, and scale it with the square root of the stellar luminosity. Exo-Earth orbits are assumed circular, based on the eccentricity-radius correlation from \citet{Kane2012}. We assume $\eta_{\oplus}=0.1$, the fraction of stars with Earth-sized planets in the OKHZ, based on planet occurrence estimates ranging from 0.05 to 0.2 for Sun-like stars \citep{Petigura2013,Marcy2014,Silburt2015}. In this simulation, we also include different amounts of exozodiacal dust in units of zodis, using the ''zodi'' definition from \citet{Stark2014}.

\begin{figure}[!ht]
\centering
\resizebox{\hsize}{!}{
\includegraphics{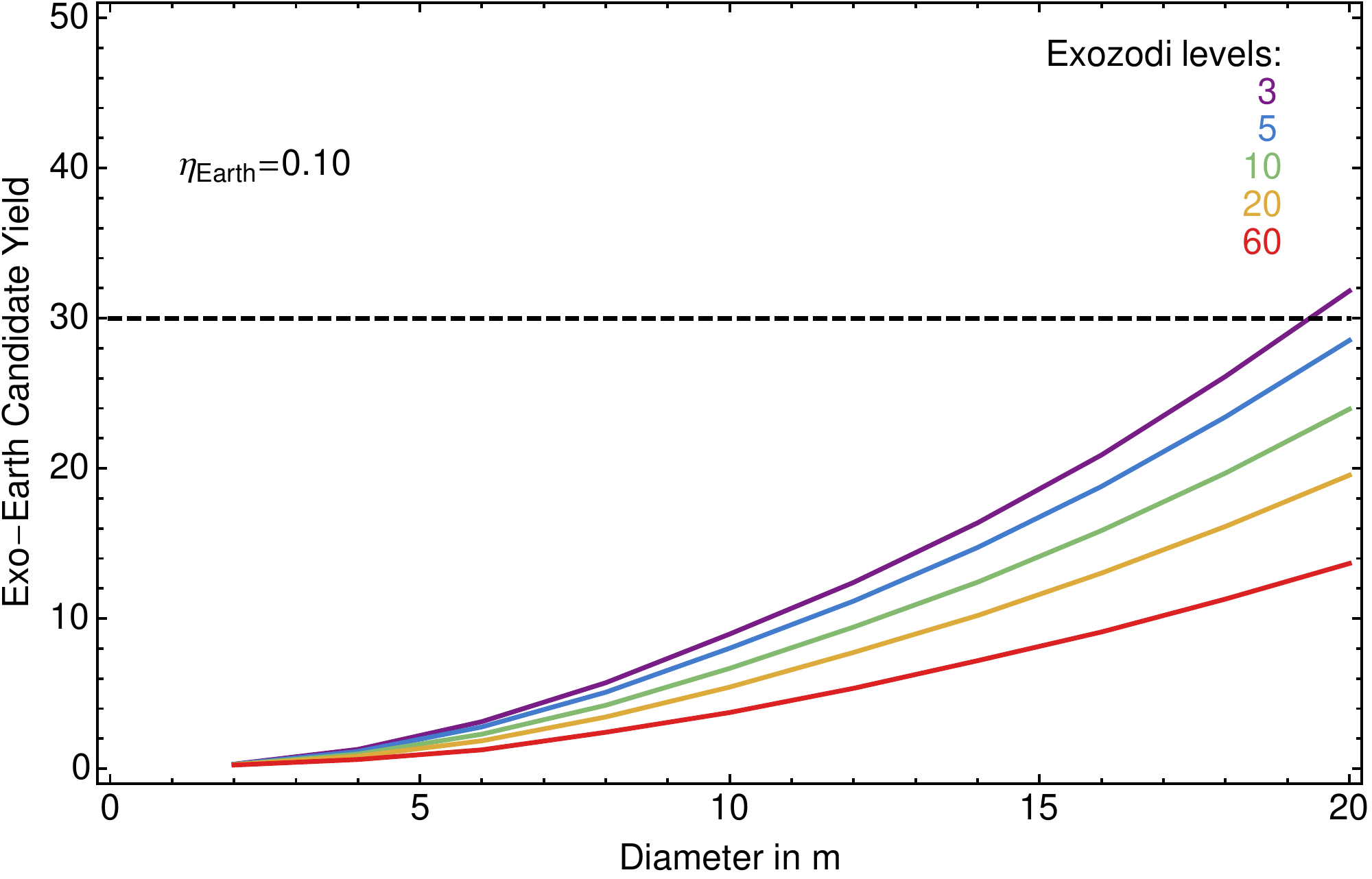}
}
\caption{ExoEarth candidate Yield with our coronagraph design as a function of the telescope diameter for different exozodi levels. Following the 12\,m telescope scenario for the ATLAST concept, we find 12.5 detectable exo-Earth candidates within 3 zodi levels for this mission with our coronagraph characteristics (IWA, OWA, contrast, Airy throughput), assuming a perfect and stable wavefront. Unsurprisingly, the exoplanet number drops at 5 with 60 exo-zodi levels but the initial exo-Earth yield is recovered by increasing the aperture size up to 20\,m.}
\label{fig:yield_estimate}
\end{figure}

Figure \ref{fig:yield_estimate} represents the yield of exo-Earth candidates as a function of the telescope diameter for different exozodi levels. The yield for a mission equipped with our coronagraph improves substantially as the telescope diameter increases. Assuming a future constraint on the median exozodi level of $<$3 zodis, we estimate our coronagraph design with a 12\,m segmented aperture can detect 12.5 exo-Earth candidates, assuming a perfect and stable wavefront.

Clearly, our novel design here produces a moderate yield for such large telescope. However, we have here presented a $10^{10}$ raw contrast design as a first iteration, representing a proof of existence of coronagraphs that can both be manufactured as of today and handle large segmented aperture in terms of starlight attenuation for Earth twin observations. This preliminary design leaves room for yield improvement in coronagraphy, in particular with the APLC/SP hybrid concept, by exploring trade-offs in coronagraph design parameters (focal plane mask size and type, inner and outer edges of the dark zone, Lyot stop geometry). 

In an attempt to identify the key points for yield enhancement, we assess the impact of the coronagraph parameters on the yield, see Table \ref{table:yield}. A few observations can be derived from its analysis:
\begin{enumerate}
\item the dark region outer edge is a good parameter to degrade in coronagraph design since it barely alters the exo-Earth yield and its reduction allow increasing the amount of flux in the PSF core. In contrast, maintaining a large dark region outer edge would favor the detection of companions, such as mature Jupiter-like planets at large separations
\item trading off contrast for IWA is worthwhile if we gain a $>0.5\,\lambda_0/D$ in IWA for a factor of 2 degradation in contrast
\item the PSF core size impacts our design performance
\item since we estimate contrast with respect to the PSF peak in the absence of focal plane mask, the throughput does not affect the amount of leaked starlight hitting the detector. Thus, increasing the throughput is more valuable than enlarging the bandwidth or mission lifetime because it does not increase one of the sources of noise.
\end{enumerate}
These remarks give us a clear path for further coronagraph designs to enhance the yield and provide a maximized science return with future large space observatories.  

\begin{table}[!ht]
\caption{Impact of coronagraph parameters on Exo-Earth yield, assuming a 10\,m telescope with 3 zodis of exozodi. In this parametric study, the apodizer has not been re-optimized for each set of parameters.}
\centering
\begin{tabular}{c c c}
\hline\hline
Parameter & Change & Exo-Earths number change\\
\hline
\multirow{2}{*}{IWA} & -0.5\,$\lambda_0/D$ & +1.1\\
 & +0.5\,$\lambda_0/D$ & -1.0\\
\multirow{2}{*}{OWA} & -2.0\,$\lambda_0/D$ & -0.1\\
 & +2.0\,$\lambda_0/D$ & +0.1\\
\multirow{2}{*}{Contrast} & x 0.5 & -1.7\\
 & x 2.0 & +1.0\\
\multirow{2}{*}{Bandwidth} & x 0.5 & -2.7\\
 & x 2.0 & +3.3\\
\multirow{2}{*}{Throughput} & x 0.5 & -4.0\\
 & x 2.0 & +4.7\\
\multirow{2}{*}{PSF radial size} & x 0.5 & +6.4\\
 & x 2.0 & -4.7\\
\multirow{2}{*}{Total exposure time} & x 0.5 & -2.7\\
 & x 2.0 & +3.3\\
\hline
\end{tabular}
\label{table:yield}
\end{table}

\section{Conclusion}\label{sec:conclusion}
We have proposed new APLC/shaped pupil hybrid solutions suitable for arbitrary telescope apertures including central obstruction, segmentation, and spiders. These new solutions are particularly interesting since they are already manufacturable, by building directly on shaped-pupil technology. 

To illustrate our new APLC solutions, we have presented a diffraction suppression system that could handle on-axis telescope with pupil features for broadband observations with $10^{10}$ contrast and that is virtual insensitivity to low-order aberrations and star diameter. In this example, a maximum Airy throughput of 18\% is achieved to detect the photons from off-axis companions through our coronagraph design. 

We have presented our approach with an APLC solution as a simple illustration of the concept. Further improvements are expected with an exploration of the parameter space that includes FPM diameter, Lyot stop feature sizes, dark zone dimensions, and spectral bandwidth to find coronagraph solutions with a maximized number of off-axis companion photons for a given integration time. Also, we have not explored yet different telescope geometries that still have degrees of freedom e.g. obstruction ratio, width of support structures. We can also imagine an optimal combination of our hybrid APLC/SP solutions with the Active Combination of Aperture Discontinuities \citep[ACAD,][]{Pueyo2013,Mazoyer2015} for diffraction control with arbitrary apertures in particular as a way to improve the off-axis companion throughput. Our approach presented in the context of APLC can be applied to vector vortex coronagraphs \citep{Mawet2005} and dual-zone phase mask coronagraphs \citep{Soummer2003b,N'Diaye2012a}, or other complex masks \citep{Guyon2010b} to provide solutions with $10^{10}$ contrast at very small IWA ($<$ 2\,$\lambda_0/D$).

A comparison of the derived solutions in terms of exo-Earth yield estimate can be made to find the coronagraph design that maximizes the number of observable targets for a given mission. Such studies will be helpful to prepare future Design Reference Missions.

\acknowledgments
The authors are grateful to the referee for the insightful suggestions and comments on the manuscript. This work is supported by the National Aeronautics and Space Administration under Grant NNX12AG05G and NNX14AD33G issued through the Astrophysics Research and Analysis (APRA) program (PI: R. Soummer). This material is also partially based upon work carried out under subcontract 1496556 with the Jet Propulsion Laboratory funded by NASA and administered by the California Institute of Technology. The authors would like to thank A. J. E. Riggs, N. Zimmerman, N. J. Kasdin, and R. Vanderbei for helpful discussions on shaped pupil designs and modeling. The authors also warmly acknowledge Aki Roberge and his team for Haystacks model of the solar system. 

\bibliographystyle{apj}   
\bibliography{biblio_mam}   

\end{document}